\documentclass[multphys,vecphys]{svmult}

\usepackage{makeidx}         
\usepackage{graphicx}        
\usepackage{multicol}        
\usepackage[bottom]{footmisc}

\makeindex             

\begin{document}

\title*{Metal-rich Globular Clusters: An
 Unaccounted Factor Responsible for Their Formation?}

\author{V. V. Kravtsov\inst{1,2}}

\institute{Instituto de Astronom\'ia, Universidad Cat\'olica del Norte,
              Avenida Angamos 0610, Casilla 1280, Antofagasta, Chile
\texttt{vkravtsov@ucn.cl}
\and Sternberg Astronomical Institute, University Avenue 13,
              119899 Moscow, Russia}

\maketitle

{\bf Abstract.} Presently unaccounted  but quite probable "chemical factor" 
may be responsible for the formation of old metal-rich globular clusters 
(MRGCs) in spheroids, as well as of their conterparts, young (intermediate--age) 
massive star clusters (MSCs) in irregulars. Their formation presumably occurs 
$\sim$ at the same stage of the host galaxies' chemical evolution and is 
related to the essentially increased SF activity in the hosts around the same 
metallicity, $\sim$Z$\odot$/3 ([Fe/H]$\sim-0.5$). It is achieved very 
soon in massive spheroids, later in  lower-mass spheroids, and (much) more later in 
irregulars.

\section{MSCs as Young Counterparts of Old MRGCs}
\label{sec:1}
Are merger of gas--rich spirals and multiphase collapse the only contributors to the 
formation of old MRGC populations? I argue that MSCs (compact populous 
and super--star clusters with M$\geq 10^4M_\odot$) in the LMC and other irregulars are 
counterparts of the old MRGCs and that another reason (quantitative 
and qualitative changes of the dust?) leads to (favors) their formation.

Peak metallicities of MRGCs in early--type galaxies with stellar masses differing by nearly 
2.5 order of magnitude are estimated by \cite{peng} to fall between 
$-0.7\leq$[Fe/H]$\leq-0.2$. The MRGC populations are assumed to be coeval, and their color 
trend 
is fully attributed to their metallicity trend. However, this is not supported by data on 
timing of spheroids' formation: the more massive spheroid, the shorter timescale of its 
formation (\cite{granato}, \cite{thomas}, \cite{caldwell}). 
Real scatter of the MRGC peak metallicities around mean, [Fe/H]$\sim-0.5$, may be 
at least twice as lower, by accepting conservative estimate of possible systematic 
age difference of $\sim$5 Gyr between MRGCs in spheroids of the range of mass.

\begin{figure}
  \centering
  \includegraphics[angle=-90,width=3.6cm]{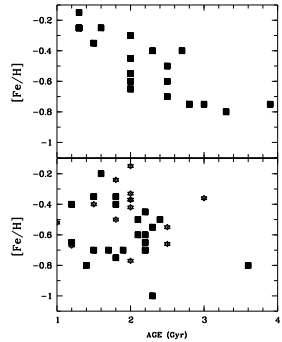}
   \caption{Upper panel: the age--metallicity relation for the LMC
intermediate--age field stars; lower panel: the same for the LMC
intermediate--age populous star clusters.}
         \label{agemet}
\end{figure}

According  to  \cite{geisler}, a mean metallicity of the populous star clusters formed in the 
LMC 1--3 Gyr ago is close to [Fe/H ]= -0.5, irrespective of their age and location in the 
galaxy. However, metallicity of the field stars near these clusters exhibits obvious 
dependence on age (see Fig.~\ref{agemet}, where  squares  and asterisks  are data from  
\cite{piatti} and \cite{olszewski}, respectively).  Moreover, the MDF for the disk stars of 
the LMC is virtually identical with that for the old red giant stars in the halo of NGC 5128 
and M31 (\cite{harris}), reaching its maximum somewhere near [Fe/H]$\sim-0.5$. 
Published data on generic (mean) metallicities of populations of MSCs and/or on their 
hosts in irregulars with increased (bursting) SF 
activity (NGC1140, NGC1156, NGC1313, NGC1569, NGC1705, NGC4214, NGC4449, NGC5253, NGC6745, 
and IC 10) show that the metallicities fall, as those of MRGCs do, around $\sim$Z$\odot$/3, 
between 0.004$\leq$Z$\leq$0.008 ($-0.7\leq$[Fe/H]$\leq-0.3$). For details and references, see 
\cite{kravtsov}.

\section{Implications}
\label{sec:2}
Both the most probable formation of GCs (MSCs) and internally regulated SF activity 
increasing in the hosts near $\sim$Z$\odot$/3 may shed more light on: the same 
metallicity value of the intracluster gas in galaxy clusters; starburst phenomenon in 
isolated galaxies; formation of MSCs in the disks of isolated spirals, etc. The 
difference between the age-metallicity relations for MSCs and stars in the LMC implies 
(provided it is the same in spheroids) different concentrations of MRGCs and 
stars to the centers of spheroids even under negligible GC disruption in the central parts 
and no merger of gas-rich spirals.

\printindex
\end{document}